\documentclass[aps,reprint,prd,showpacs,amsmath,amssymb,nofootinbib]{revtex4-1}
\usepackage{graphicx}
\usepackage{color}
\usepackage{times}
\usepackage{inputenc}
\usepackage{bm}
\usepackage{multirow}
\usepackage{float}
\usepackage{url}
\usepackage{natbib}
\usepackage{tensor}
\usepackage[colorlinks=true,citecolor=blue,urlcolor=blue,linkcolor=red]{hyperref}
\usepackage[normalem]{ulem}
\usepackage{lineno}

\def\msun{\,M_{\odot}}
\def\fm3{\;\text{fm}^{-3}}
\def\mev{\;\text{MeV}}

\begin{document}


\title{Bayesian inference of strangeon matter using the measurements of PSR J0437-4715 and GW190814}
\author{Wen-Li Yuan$^{1}$}
\email{wlyuan@pku.edu.cn} 
\author{Chun Huang$^{2}$}
\email{chun.h@wustl.edu}
\author{Chen Zhang$^{3}$} 
\email{iasczhang@ust.hk} 
\author{Enping Zhou$^{4}$}
\email{ezhou@hust.edu.cn} 
\author{Renxin Xu$^{1}$}
\email{r.x.xu@pku.edu.cn} 
\affiliation{$^1$School of Physics and State Key Laboratory of Nuclear Physics and Technology, Peking University, Beijing 100871, China;\\ 
}
\affiliation{$^2$Physics Department and McDonnell Center for the Space Sciences, Washington University in St. Louis; MO, 63130, USA;\\
}
\affiliation{$^3$The HKUST Jockey Club Institute for Advanced Study, The Hong Kong University of Science and Technology, Hong Kong, China;\\
}
\affiliation{$^4$Department of Astronomy, School of Physics, Huazhong University of Science and Technology, Wuhan 430074, China\\
}

\date{\today}

\begin{abstract}
The observations of compact star inspirals from LIGO/Virgo, combined with mass and radius measurements from NICER, provide a valuable tool to study the highly uncertain equation of state (EOS) of dense matter at the densities characteristic of compact stars. In this work, we use a Bayesian statistical method to constrain the solid states of strange-cluster matter, called strangeon matter, as the putative basic units of the ground state of bulk strong matter, incorporating the mass and radius measurements of PSR J0030+0451, PSR J0740+6620, and the recent data for the $1.4\msun$ pulsar PSR J0437-4715. We also include constraints from gravitational wave events GW170817 and GW190814. Under the prior assumption of a finite number of quarks in a strangeon, $N_{\rm q}$, our analysis reveals that current mass-radius measurements favor a larger $N_{\rm q}$. Specifically, the results support the scenario where a strangeon forms a stable bound state with $N_{\rm q}=18$, symmetric in color, flavor, and spin spaces, with a relatively strong Bayesian evidence. The comparative analyses of the posterior EOS parameter spaces derived from the three-parameter model and two-parameter model demonstrate a consistent prediction under identical observational constraints. In particular, our results reveal that the most probable values of the maximum mass are found to be $3.58^{+0.16}_{-0.12}\msun$ ($3.65^{+0.18}_{-0.16}\msun$) at the $90\%$ confidence level for three-parameter (two-parameter) EOS, based on the joint analysis of PSR J0030+0451, PSR J0740+6620, PSR J0437-4715, GW170817, and GW190814. Correspondingly, the radii for $1.4\msun$ and $2.1\msun$ stars are $12.04^{+0.27}_{-0.31}~\rm km$ ($12.16^{+0.26}_{-0.31}~\rm km$) and $13.43^{+0.31}_{-0.32}~\rm km$ ($13.60^{+0.29}_{-0.34}~\rm km$), respectively. The tidal deformability $\Lambda_{\rm 1.4}$ for a $1.4\msun$ star is $205^{+32.49}_{-32.53}$ ($212.36^{+53.71}_{-32.66}$). These results may impact the research of multiquark states, which could potentially improve our understanding of the nonperturbative strong interaction.

\end{abstract}
\maketitle 

\section{Introduction} 
The equation of state (EOS) of dense quantum chromodynamics (QCD) matter has been the subject of extensive studies during the last few decades,  which provides information on the internal structure and composition of compact stars~\cite{1999LNP...516..162M,2005PrPNP..54..193W,2017RvMP...89a5007O,2018RPPh...81e6902B,2019PrPNP.10903714B,2022PhRvL.128t2701K}.
Thanks to the increasing number of electromagnetic (EM) observations, such as radio and X-ray, as well as gravitational wave detections, the ever-increasing data from nuclear physics experiments and astrophysical observations have provided valuable information on the EOS of such objects~\cite{1999LNP...516..162M,2005PrPNP..54..193W,2017RvMP...89a5007O,2018RPPh...81e6902B,2019PrPNP.10903714B}. The discoveries of a few pulsars over $2\msun$ have put stringent constraints on the EOS of supranuclear matter~\cite{2021ApJ...918L..28M,2021ApJ...918L..27R}. It requires that the matter inside such compact stars must be stiff enough to sustain these massive stable configurations. Conversely, the measurement of tidal deformability from the GW170817 event suggests smaller radii for the low-mass compact stars~\cite{2017PhRvL.119p1101A,2018PhRvL.121p1101A}, implying that the EOS becomes softer at densities associated with the low-mass configurations. 
In addition to the results of the masses and radii for PSR J0740+6620~\cite{2021ApJ...918L..28M,2021ApJ...918L..27R} and PSR J0030+0451~\cite{2019ApJ...887L..24M,2019ApJ...887L..21R,2024ApJ...961...62V}, the recent new result on the
radius measurement for the brightest rotation-powered millisecond X-ray pulsar PSR J0437-4715~\cite{2024ApJ...971L..20C,2024ApJ...971L..19R} with $\sim 1.4\msun$ reports a radius very close to that from the gravitational wave observation of the binary neutron star merger event GW170817. These have led to a steady improvement in our understanding of dense matter EOS.  

According to the Bodmer-Witten hypothesis~\cite{1971PhRvD...4.1601B,1984PhRvD..30..272W}, compact stars could be formed by self-bound deconfined quarks that make up the entire star, effectively a quark star. The component inside the self-bound quark stars depends on what is the true ground state of the baryonic matter. After decades of speculation, strange quark stars composed of strange quark matter~\cite{1998PhLB..438..123D,1989PhLB..229..112C,1993PhRvD..48.1409C,1999PhLB..457..261B,1999PhRvC..61a5201P,2005PhRvC..72a5204W,2010MNRAS.402.2715L,2018PhRvD..97h3015Z,2018PhRvD..98h3013L,2019PhRvD..99j3017X,2000PhRvC..62b5801P,2021EPJC...81..612B,2021ChPhC..45e5104X,2022PhRvD.105l3004Y,2024PhRvD.110j3012Z,2024FrASS..1109463Z} 
and up-down quark stars with up-down quark matter inside~\cite{2018PhRvL.120v2001H,2019PhRvD.100d3018Z,2020PhRvD.102h3003R,2020PhRvD.101d3003Z,2021PhRvD.103f3018Z,2024Ap&SS.369...29S} are both alternative physical models for neutron stars. It is also intriguing that, besides the new degree of freedom of strangeness, the {\em non-perturbative} QCD is, nevertheless, worth noting, which could led quarks to be localized in clusters. This strange cluster~\cite{2003ApJ...596L..59X} has been renamed strangeon, a nucleon-like bound state in fact. The strangeon matter has intrinsically stiff EOSs~\cite{2003ApJ...596L..59X,2006MNRAS.373L..85X,2008MNRAS.384.1034P,2009MNRAS.398L..31L,2012MNRAS.424.2994L,2014MNRAS.443.2705Z,2016ChPhC..40i5102L,2018RAA....18...24L,2019EPJA...55...60L,2021RAA....21..250L,2022MNRAS.509.2758G,2022IJMPE..3150037M,2023PhRvD.108f3002Z,2023PhRvD.108l3031Z} and has been proposed to support massive pulsars ($>2\msun$) prior to the announcement of the first massive pulsar PSR J1614-2230~\cite{2010Natur.467.1081D}.  And it potentially supports the GW190814 secondary object~\cite{2020ApJ...896L..44A}, which falls into the so-called “mass-gap” category, to be a strangeon star.

However, due to the non-perturbative difficulties from the first-principle QCD, the description of strangeon matter EOS could be derived phenomenologically from the Lennard-Jones potential model,~\cite{1924RSPSA.106..441J} which is characterized by three parameters, $\epsilon$, $n_{\rm sur}$ and $N_{\rm q}$, representing the depth of the potential wall, the surface baryon number density of strangeon star, and the number of quarks in one strangeon, respectively. Nevertheless, the specific details of these physical parameters are not clear enough. Due to the ever-increasing data from astronomical observations, we have the opportunity to perform a systematic analysis of the strangeon matter EOS utilizing the robust Bayesian statistical method. This approach facilitates the inference of posterior distributions for a number of physical parameters by integrating a set of measured data, thereby refining our understanding. See Refs.~\cite{2010ApJ...722...33S,2019MNRAS.485.5363G,2019ApJ...887L..22R,2020ApJ...893L..21R,2020ApJ...897..165T,2021PhRvC.103c5802X,2023PhRvD.108d3002T,2024MNRAS.529.4650H,2024PhRvD.110f3040M,2024PhLB..85939128G,2024arXiv241014572H} for neutron star EOS inference and Refs.~\cite{2020ApJ...905....9T,2021ApJ...917L..22M,2021MNRAS.506.5916L,2024PhRvD.109d3054D,2024arXiv240911103W} for quark star EOS inference as examples.
This work is the first attempt to constrain the strangeon matter EOS applying Bayesian analysis based on the astronomical observations, and to explore the mass-radius (M-R) posterior distributions for strangeon stars. In this analysis, we incorporate not only the recent simultaneous mass and radius measurements of PSR J0030+0451 from NICER~\cite{2019ApJ...887L..21R,2019ApJ...887L..24M,2024ApJ...961...62V}, the observation measurement of PSR J0740+6620~\cite{2021ApJ...918L..27R,2021ApJ...918L..28M,2021ApJ...915L..12F,2024ApJ...974..294S}, and the gravitational wave event GW170817~\cite{2017PhRvL.119p1101A,2018PhRvL.121p1101A}, but also the constraints from the recent new result of PSR J0437-4715~\cite{2024ApJ...971L..20C,2024ApJ...971L..19R} and the observed ~$2.6\msun$ compact object in the GW190814's secondary component~\cite{2020ApJ...896L..44A}, allowing for a comprehensive exploration of the strangeon matter EOS.

The paper is organized as follows. Section~\ref{Sec: EOS} is a brief overview of the strangeon matter EOS for strangeon stars, where we consider two different forms of the model in the following analysis. In Section~\ref{Sec: Bayesian}, we present the employed astronomical observations and the Bayesian inference procedure. Section~\ref{Sec: Results} discusses the results and the properties of strangeon stars, then we summarize in Section~\ref{Sec: Summary}.

\section{Strangeon matter EOS}\label{Sec: EOS}
In this section, we formulate the EOS model to describe strangeon matter, then define the two model formulations we explored in this work.

\subsection{EOS}
Following previous studies~\cite{2003ApJ...596L..59X,2006MNRAS.373L..85X,2008MNRAS.384.1034P,2009MNRAS.398L..31L,2012MNRAS.424.2994L,2014MNRAS.443.2705Z,2016ChPhC..40i5102L,2018RAA....18...24L,2019EPJA...55...60L,2021RAA....21..250L,2022MNRAS.509.2758G,2022IJMPE..3150037M,2023PhRvD.108f3002Z,2023PhRvD.108l3031Z}, the interaction potential between two strangeons is described by the Lennard-Jones potential~\cite{1924RSPSA.106..441J}:
\begin{equation}
    U(r)=4 \epsilon\left[\left(\frac{\sigma}{r}\right)^{12}-\left(\frac{\sigma}{r}\right)^6\right]\ ,
\end{equation}
where $\epsilon$ is the depth of the potential well, $r$ is the distance between two strangeons, and $\sigma$ is the distance at which $U(r)=0$, representing the characteristic separation between two strangeons where the attractive and repulsive forces exactly cancel out, resulting in a zero potential energy. A larger $\epsilon$ will then indicate a larger repulsive force at short range and thus map to a stiffer EOS.

The mass density $\rho$ and pressure density $P$ of strangeon matter at zero temperature derived from Lennard-Jones potential~\cite{2009MNRAS.398L..31L} reads
\begin{equation}
\begin{aligned}
& \rho=2 \epsilon\left(A_{12} \sigma^{12} n^5-A_6 \sigma^6 n^3\right)+n N_{\mathrm{q}} m_{\mathrm{q}} \ , \\
& P=n^2 \frac{\mathrm{d}(\rho / n)}{\mathrm{d} n}=4 \epsilon\left(2 A_{12} \sigma^{12} n^5-A_6 \sigma^6 n^3\right)\ , \label{eq:eos_three}
\end{aligned}
\end{equation}
where $A_{12}=6.2, A_6=8.4$, and $n$ is the number density of strangeons. $N_{\mathrm{q}} m_{\mathrm{q}}$ is the mass of a strangeon with $N_{\mathrm{q}}$ being the number of quarks in a strangeon and $m_{\rm q}$ being the average constituent quark mass. We take the mass of quarks to be $m_{\rm q} = 300 \mev$, which is about one-third of the nucleon’s mass. The contributions from degenerate electrons and vibrations of the lattice are neglected due to their expected smallness. 

At the surface of strangeon stars, the pressure becomes zero, and the surface number density of strangeons is $\left[A_6 /\left(2 A_{12} \sigma^6\right)\right]^{1 / 2}$. The relationship between the baryon number density $n_{\rm b}$ and strangeon number density $n$, given by $n_{\rm b}=n N_{\rm q}/3$, leads to the surface baryon number density $n_{\rm sur}$ being expressed as $\left(A_6/2 A_{12}\right)^{1 / 2}N_{\rm q}/3\sigma^3$. Accordingly, the EOS can be rewritten into a form that depends on the three-parameter set $\left(N_{\rm q}, \epsilon, n_{\rm sur} \right)$:
\begin{equation}
\begin{aligned}
& \rho=\frac{1}{9} \epsilon \frac{A_6^2}{A_{12}}\left(\frac{N_{\rm q}{ }^4}{18 n_{\rm sur}^4} n^5-\frac{N_{\rm q}^2}{n_{\rm sur}^2} n^3\right)+m_{\rm q} N_{\rm q} n \ , \\
& P=\frac{2}{9} \epsilon \frac{A_6^2}{A_{12}}\left(\frac{N_{\rm q}^4}{9 n_{\rm sur}^4} n^5-\frac{N_{\rm q}^2}{n_{\rm sur}^2} n^3\right)\ .
\end{aligned}
\end{equation}

\begin{figure}
\centering
{\includegraphics[width=0.45\textwidth]{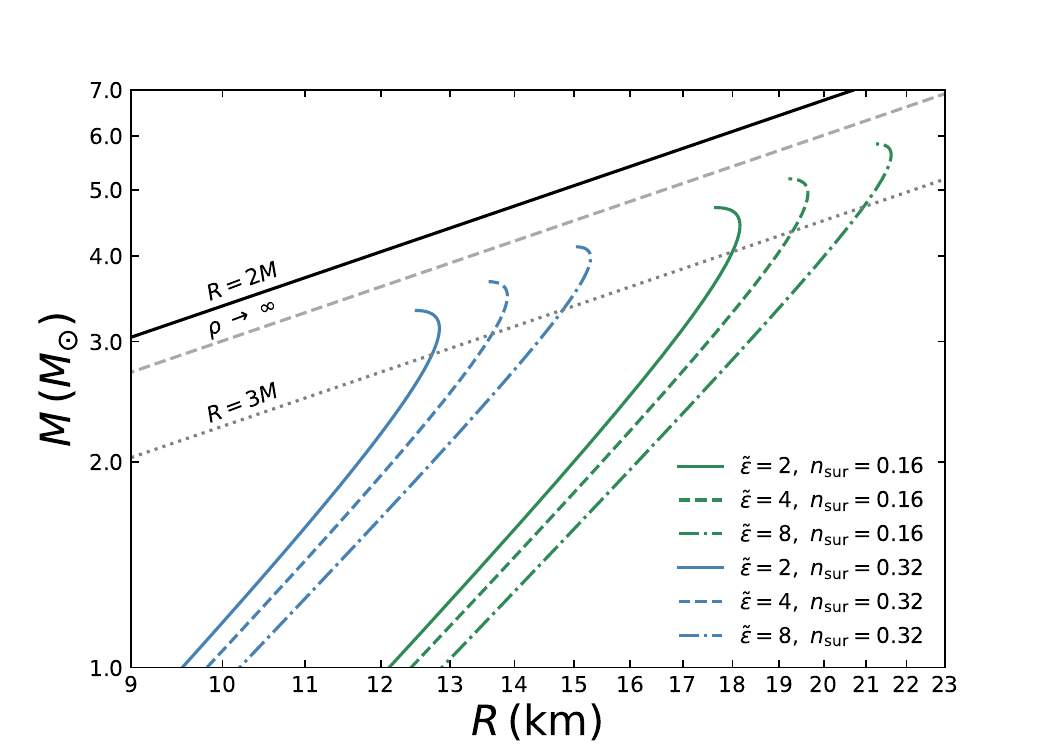}}
\caption{The M-R relations within the two-parameter model for various parameter sets of $\tilde{\epsilon}$ and $n_{\rm sur}$ in units of$\mev$ and$\fm3$, respectively. The coordinate scale is logarithmic. The black hole limit $R=2M$, and the limit for central pressure to be infinite $R=9/4M$ are also shown together.}  
\label{fig:eos_mr} 
\end{figure}

By defining $\tilde{\epsilon}=\epsilon / N_{\rm q}$ and $\bar{n}=N_{\rm q} n / n_{\rm sur}$, the simpler form of strangeon matter EOS can be derived as follows~\cite{2023PhRvD.108l3031Z}: 
\begin{equation}
\begin{aligned}
& \frac{\rho}{n_{\rm sur}}=\frac{a}{9} \tilde{\epsilon}\left(\frac{1}{18} \bar{n}^5-\bar{n}^3\right)+m_{\rm q} \bar{n}\ , \\
& \frac{P}{n_{\rm sur}}=\frac{2 a}{9} \tilde{\epsilon}\left(\frac{1}{9} \bar{n}^5-\bar{n}^3\right) \ , \label{eq:eos_two}
\end{aligned}
\end{equation}
with $a=A_6^2 / A_{12}$.  Note that $\bar{n}=3$ at star surface where $P=0$. In the following, for convenience, we define the model containing $N_{\rm q}$, $\epsilon$, $n_{\rm sur}$  as free parameters to be the three-parameter model, and $\tilde{\epsilon}$, $n_{\rm sur}$ to be the two-parameter model, where $\tilde{\epsilon}$ means the depth per quark of the potential wall in one strangeon.

In Fig.~\ref{fig:eos_mr}, we display M-R relations of strangeon matter EOS with different sets of parameters $\tilde{\epsilon}$ and $n_{\rm sur}$ within the two-parameter model, while the three-parameter EOS are discussed in Ref.~\cite{2022MNRAS.509.2758G}. To better illustrate the approximately linear relationship between mass and radius at low densities, we plot the M-R relations in logarithmic space. The results indicate that the ratio of $\epsilon $ to $ N_{\rm q}$, denoted as $\tilde{\epsilon}$, is the key factor influencing the shape of M-R curve, as shown in Eq.~\ref{eq:eos_two}.  
In other words, changing the values of $\epsilon$ and $N_{\rm q}$ while keeping the $\tilde{\epsilon}$ constant does not impact the M-R relations. Increasing $n_{\rm sur}$ significantly changes the surface energy density as well as the whole range of energy densities, resulting in a softer EOS, and hence a lower maximum mass. 

\begin{figure*}
\centering
{\includegraphics[width=0.45\textwidth]{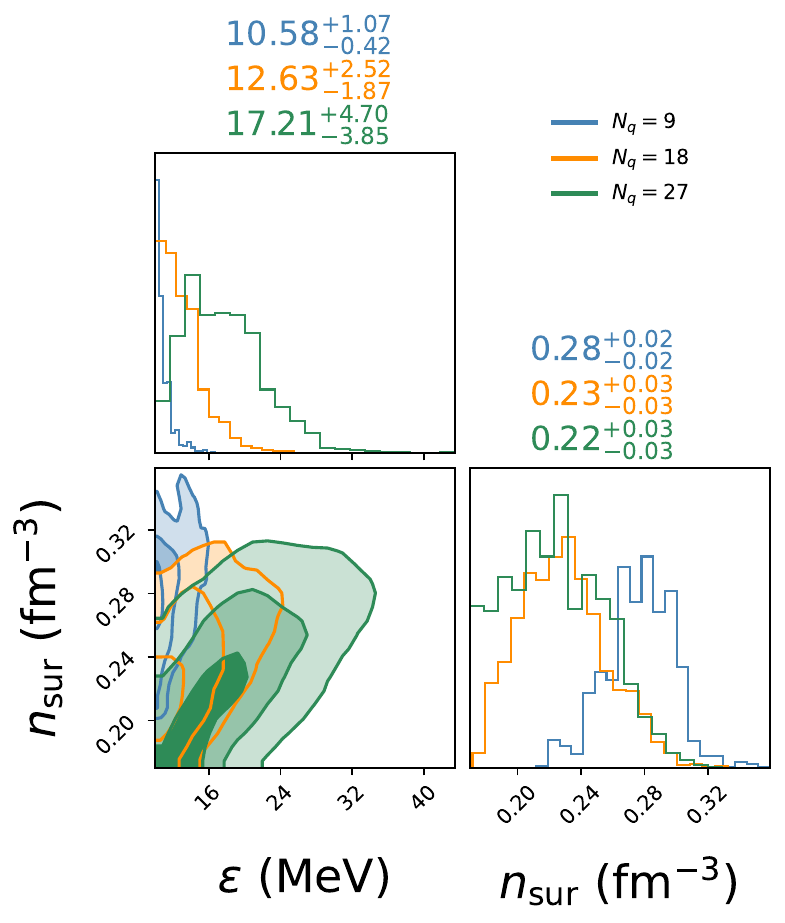}}
{\includegraphics[width=0.45\textwidth]{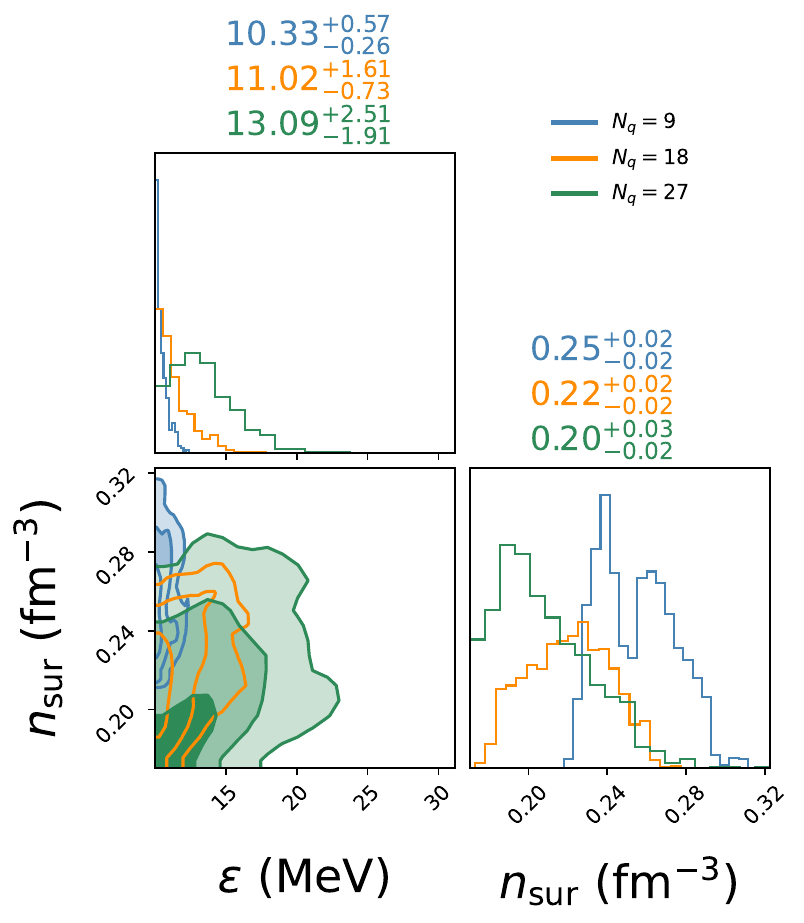}}
\caption{Left panel: The posterior distribution of the model parameters under the constraints of PSR J0030 + 0451 and PSR J0740 + 6620 at different cases of fixed $N_{\rm q}$. The contour levels in the corner plot correspond to the 68.3$\%$, 95.4$\%$, and 99.7$\%$ confidence levels, going from dark to light.
Right panel: The posterior distribution of the joint analysis with PSR J0030 + 0451, PSR J0740 + 6620, and the new result measurement of PSR J0437-4715. The confidence levels are the same as in the left panel.}  
\label{fig:posterior_NICER}
\end{figure*}
\begin{figure*}
\centering
{\includegraphics[width=0.49\textwidth]{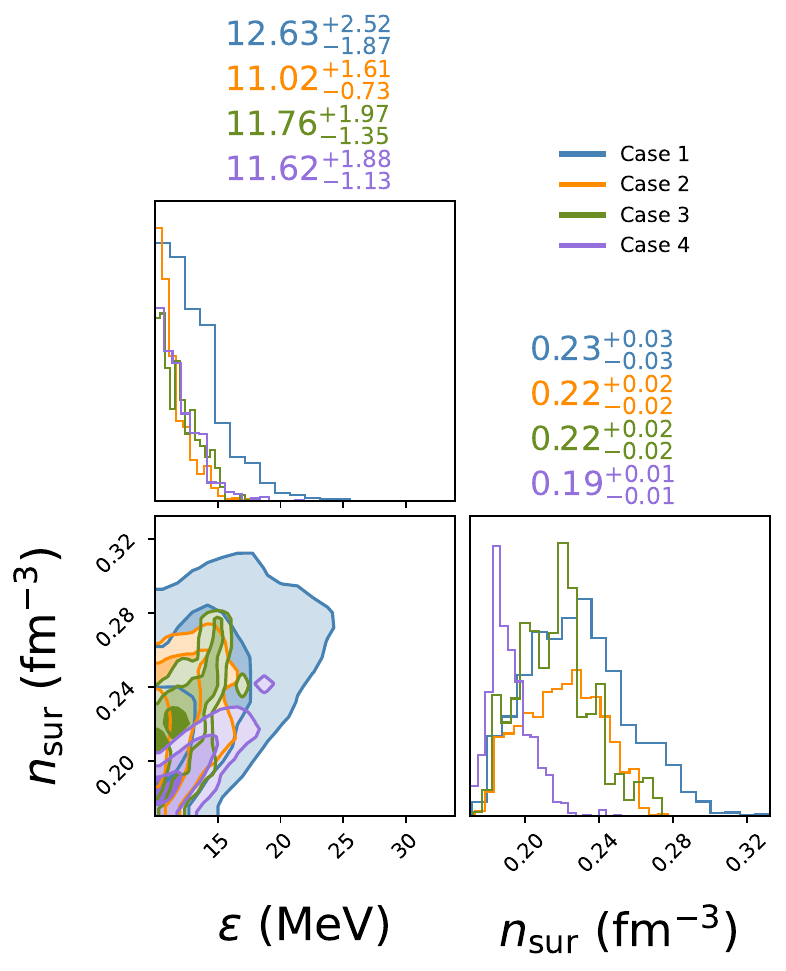}}
{\includegraphics[width=0.49\textwidth]{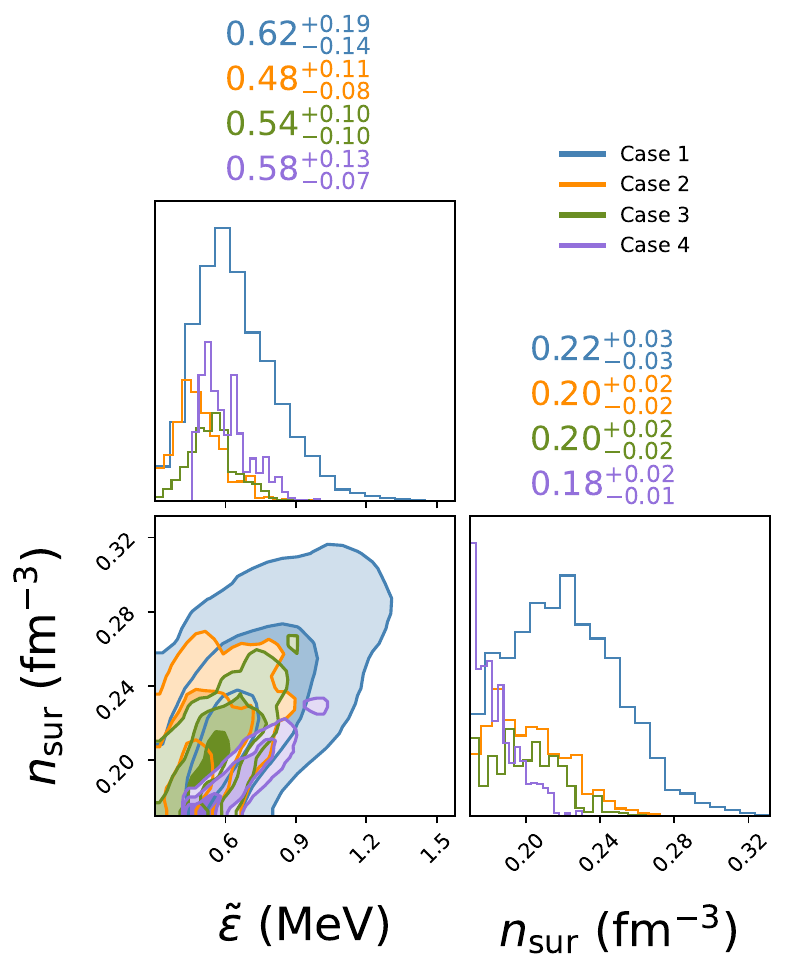}}
\caption{The posterior distributions of the three-parameter model EOS at fixed $N_{\rm q}=18$ and two-parameter model EOSs are shown in the left and right panels, respectively. Each distribution indicates the different constraints of Case 1, Case 2, Case 3, and Case 4. The contour levels in the corner plot correspond to the 68.3$\%$, 95.4$\%$, and 99.7$\%$ confidence levels, going from dark to light.
}   
\label{fig:compare_three_and_two_parameter}
\end{figure*}

\begin{table*}
\centering
\caption{The Bayesian evidence with and without PSR J0437-4715 for different $N_{\rm q}$. The $N_{\rm q}^{*}$ denotes the relatively supported EOS parameter by comparing the Bayes factor with the analysis under identical observational data, which will be discussed in detail below.
}
                  \vskip+2mm
\renewcommand\arraystretch{1.5}
\begin{ruledtabular}
\begin{tabular*}{\hsize}{@{}@{\extracolsep{\fill}}lcccccc@{}}
 Evidence &  
 three-parameter model  & $N_{\rm q}=9$  & $N_{\rm q}^{*}=18$ & $N_{\rm q}=24$ & $N_{\rm q}=27$  &\\
\hline log\ Z  & without PSR J0437-4715 & -34.4 &  -29.8 &  -28.9  & -29.0 &\\
\hline log\ Z & with PSR J0437-4715   & -48.1 & -36.7 & -37.5 & -36.0   &\\
\end{tabular*}
\end{ruledtabular}
    \vspace{-0.4cm}
\label{table:bayes_factor}
\end{table*} 

\section{Constraints and Bayesian analysis}\label{Sec: Bayesian}
In the following, we consider the strangeon matter constituting the stars within two model formulations, and employ the Bayesian analysis to infer the posterior of strangeon matter EOS parameters, $(N_{\rm q}, \epsilon, n_{\rm sur})$ or $(\tilde{\epsilon}, n_{\rm sur})$, by applying multi-messenger observational constraints. Subsequently, we infer the allowed M-R space of strangeon stars, filtered by the observations we implied. Using this Bayesian technique, we aim to identify the value of $N_{\rm q}$ that best aligns with current observational data, and to assess whether the results obtained from three-parameter and two-parameter models are consistent under the same set of observational constraints.

\subsection{Choice of priors for model parameters}
The three-parameter model is characterized by three free parameters: $N_{\rm q}$, $\epsilon$, and $n_{\rm sur}$, which capture the unique properties of the strong interactions between strangeons as mentioned before. 
Although the exact values of these parameters remain uncertain, reasonable ranges can be inferred based on the current understanding of strong interactions. In this analysis, we assume the parameter $N_{\rm q}$ takes values from the set $N_{\rm q} = 9, 18, 21, 24, 27$, each a multiple of three, to satisfy the color neutrality requirement. Motivated by the evidence of the unstable H-dibaryon~\cite{1997NuPhA.625..167B,2000NuPhS..83..218W}, which consists of six quarks in a flavor-singlet state, we consider $N_{\rm q} \geq 9$ as a minimum for the number of quarks in a strangeon. In particular, an 18-quark strangeon is called a quark-alpha~\cite{1988PhRvL..60..677M,1991NuPhS..24...33C}, which is fully symmetric in spin, flavor, and color space, and wherein a colorless triplet of each of the spin-paired quarks can sit in an S-state. While the theoretical upper limit of $N_{\rm q}$ is currently unknown, we set a modestly higher upper bound of $N_{\rm q} = 27$. As the results of the following Bayesian inference will show in Table~\ref{table:bayes_factor}, $N_{\rm q} = 27$ provides an adequate prior, as the Bayes factor comparison shows no substantial improvement over the $N_{\rm q} = 18$ case. Consequently, we adopt $N_{\rm q} = 27$ as the upper limit in this Bayesian framework. 
The nucleon-nucleon scattering data indicate that the inter-nucleon potential well lies in the range of $\sim 50 -120 \mev$ for the ${}^{1}\text{S}_{0}$ (spin-singlet and S-wave) channel~\cite{1994PhRvC..49.2950S,1995PhRvC..51...38W,2001PhRvC..63b4001M}. Since the strong interactions are not sensitive to the flavor of quarks, in this Bayesian analysis we choose $\epsilon$ spanning in the range of $10-170\mev$. The surface baryonic density $n_{\rm sur}$ should be in the same order as the nuclear saturation density, $n_0=0.16\fm3$, due to the self-bound property of strangeon stars. The interactions may group the quarks more compactly compared to nuclei containing the same number of quarks. Therefore, we let $n_{\rm sur}$ lie in the range of $0.17-0.36\fm3$, which corresponds to $\sim 1\ n_0\ -2.25\  n_0$. Accordingly, the choice of the parameter $\tilde{\epsilon}$ is set in the range of $0.3-3.0 \mev$ based on our prior choices for $\epsilon$ and $N_{\rm q}$ separately for the two-parameter model. These choices are shown in Table~\ref{table:prior_posteroor_para}. Due to the lack of terrestrial experimental constraints, all parameters are investigated with uniform contributions in this work.

\subsection{Inference framework}
According to the Bayes’ theorem,  the posterior distribution of a set of model parameters $\theta$, given the observational data set $\boldsymbol{d}$ for a model $\mathcal{M}$, can be
  \begin{equation}
    p\left(\boldsymbol{\theta} \mid \boldsymbol{d}, \mathcal{M}\right)=\frac{p\left(\boldsymbol{d} \mid \boldsymbol{\theta}, \mathcal{M}\right) p\left(\boldsymbol{\theta} \mid \mathcal{M}\right)}{p(\boldsymbol{d} \mid \mathcal{M})}\ , 
\end{equation}
where $p\left(\boldsymbol{\theta} \mid \mathcal{M}\right)$ is the prior probability of the parameter set $\theta$. $ p\left(\boldsymbol{d} \mid \boldsymbol{\theta}, \mathcal{M}\right)$ is the likelihood function of the data given the model, and $p(\boldsymbol{d} \mid \mathcal{M})$ is known as evidence for the model. For a given data set, $p(\boldsymbol{d} \mid \mathcal{M})$ is a constant and can be treated as a normalization factor. 
Since different central energy densities correspond to different masses and radii, for the present analyses, we need the parameter $\varepsilon_{\rm c}$ to perform the Bayesian analyses. 
Hence, the posterior distributions of the EOS model parameters $\theta$ and center energy densities $\varepsilon_{\rm c}$ can be written as:
 \begin{equation}
    p(\boldsymbol{\theta}, \varepsilon_{\rm c} \mid \boldsymbol{d}, \mathcal{M}) \propto p(\boldsymbol{\theta} \mid \mathcal{M}) p(\varepsilon_{\rm c} \mid \boldsymbol{\theta}, \mathcal{M})
    p(\boldsymbol{d} \mid \boldsymbol{\theta}, \mathcal{M}) \ ,\label{eq:Bayesian_inference}
\end{equation}
where $p(\boldsymbol{\theta} \mid \mathcal{M})$ and $p(\varepsilon_{\rm c} \mid \boldsymbol{\theta}, \mathcal{M})$ are the prior distributions of $\theta$ and $\varepsilon_{\rm c}$ respectively. $p(\boldsymbol{d} \mid \boldsymbol{\theta}, \mathcal{M})$ is the nuisance-marginalized likelihood function (see Refs.~\cite{2021ApJ...918L..29R,2024MNRAS.529.4650H} for the detail discussions for the definition). 
The astrophysical inputs as the likelihood for our inference are explained as follows.

\subsubsection{Constraints from NICER data}
We consider the masses and radii inferred from the NICER data by Riley et al.~\cite{2019ApJ...887L..21R,2021ApJ...918L..27R,2024ApJ...961...62V}, for PSR J0030 + 0451 we use the result from Ref.~\cite{2019ApJ...887L..21R} ($M=1.34^{+0.15}_{-0.16}~\msun$ and $R=12.71^{+1.14}_{-1.19}\rm~ km$) 
and the heavy pulsar PSR J0740 + 6620 ($M=2.072^{+0.067}_{-0.066}\msun$ and $R=12.39^{+1.30}_{-0.98}\rm~km$)
by X-ray pulse profile modeling of NICER data.  Here, we also consider the impact of the new result on the mass measurement for the $\sim1.4\msun$ pulsar PSR J0437-4715~\cite{2024ApJ...971L..20C, 2024ApJ...971L..19R}. 
Using a mass prior from radio timing \cite{2024ApJ...971L..18R} people reported a mass of $M=1.418 \pm 0.037 \msun$ and a radius of $R=11.36_{-0.63}^{+0.95} \mathrm{~km}$ (68\% credible intervals) for PSR J0437-4715.

Given that all of the measurements are independent, by equating the nuisance-marginalized likelihoods to the nuisance-marginalized posterior distributions~\cite{2021ApJ...918L..29R,2024MNRAS.529.4650H}, we can rewrite the likelihood as follows:
\begin{equation}
\begin{aligned}
& p(\boldsymbol{\theta}, \varepsilon_{\rm c} \mid \boldsymbol{d}, \mathcal{M}) \propto p(\boldsymbol{\theta} \mid \mathcal{M}) p(\varepsilon_{\rm c} \mid \boldsymbol{\theta}, \mathcal{M}) \\
& \quad \times \prod_j p\left(\boldsymbol{M}_j, R_j \mid d_{\mathrm{NICER}, \mathrm{j}}\right) \ , \\ 
\end{aligned}
\end{equation}
with $j$ representing the different measurements of masses and radii inferred from the NICER data.

\begin{table*}
\centering
\caption{The most probable intervals of the EOS parameters (68.3 $\%$ confidence level) as well as the strangeon star properties ($90\%$ confidence level) in three-parameter and two-parameter models constrained by Case 1 (PSR J0030+0451 $\&$ PSR J0740+6620), Case 2 (PSR J0030+0451 $\&$ PSR J0740+6620 $\&$ PSR J0437-4715), Case 3 (PSR J0030+0451 $\&$ PSR J0740+6620 $\&$ PSR J0437-4715 $\&$ GW170817, and Case 4 (PSR J0030+0451 $\&$ PSR J0740+6620 $\&$ PSR J0437-4715 $\&$ GW170817 $\&$ GW190814), respectively. $\mathcal{U}$ means Uniform (Flat) distribution. $M_{\rm TOV}$ is the maximum mass. $R_{\rm 1.4}$ and $R_{2.1}$ are the radii of $1.4\msun$ and $2.1\msun$ stars, respectively. $\Lambda_{\rm 1.4}$ is the tidal deformability for a $1.4\msun$ star. $n_{\rm c}(n_0)_{M_{\rm TOV}}$ represents the center baryon number densities corresponding to the maximum mass $M_{\rm TOV}$, with $n_0$ being the nuclear saturation density.}
                  \vskip+2mm
\renewcommand\arraystretch{1.5}
\begin{ruledtabular}
\begin{tabular*}{\hsize}{@{}@{\extracolsep{\fill}}lcccccc@{}}
three-parameter model ($N_{\rm q}=18$) & $\mathrm{Prior} $ & Case 1  & Case 2 & Case 3 & Case 4 &\\
\hline $\epsilon ~(\rm MeV)$  & $\mathcal{U}(10, 170)$ & $12.63^{+2.52}_{-1.87}$ &  $ 11.02^{+1.61}_{-0.73}$ &  $11.76^{+1.97}_{-1.35}$  &$11.62^{+1.88}_{-1.13} $ &\\
\hline  [$\epsilon/N_{\rm q}~(\rm MeV)$  & $\ $ & $0.70^{+0.14}_{-0.10}$ &  $0.61^{+0.09}_{-0.04}$ &  $0.65^{+0.11}_{-0.08}$  &$0.65^{+0.10}_{-0.06}$]&\\
\hline $n_{\rm sur} ~(\rm fm^{-3})$ & $\mathcal{U}(0.17, 0.36)$ & $0.23^{+0.03}_{-0.03} $& $ 0.22^{+0.02}_{-0.02}$& $0.22^{+0.02}_{-0.02}$ & $0.19^{+0.01}_{-0.01}$   &\\
\hline  $R_{1.4} ~(\rm km)$ & $\ $ & $11.32^{+0.76}_{-0.61}$ &  $11.33^{+0.66}_{-0.48}$ &  $11.47^{+0.60}_{-0.51}$  & $12.04^{+0.27}_{-0.31}$ &\\
\hline  $\Lambda_{1.4} $ & $\ $ & $160.95^{+53.30}_{-45.68}$ &  $161.49^{+49.91}_{-31.94} $ &  $170.44^{+48.07}_{-40.62}$  & $205.60^{+32.49}_{-32.53} $ &\\
\hline  $R_{2.0} ~(\rm km)$ & $\ $ & $12.59^{+0.71}_{-0.81}$ &  $12.56^{+0.72}_{-0.62}$ &  $12.62^{+0.69}_{-0.60}$  & $13.28^{+0.24}_{-0.35}$ &\\
\hline  $M_{\rm TOV}(M_{\odot})$ & $\ $ & $3.51^{+0.22}_{-0.10}$ &  $3.48^{+0.21}_{-0.10}$ &  $3.51^{+0.18}_{-0.09}$  & $3.58^{+0.16}_{-0.12}$ &\\
\hline  $n_{\rm c}(n_0)_{M_{\rm TOV}}$ & $\ $ & $4.71^{+0.57}_{-0.51}$ &  $4.71^{+0.57}_{-0.42}$ &  $4.71^{+0.34}_{-0.42}$  & $4.19^{+0.01}_{-0.24}$  &\\
\hline \hline two-parameter model & $\mathrm{Prior} $ & Case 1  & Case 2 & Case 3 & Case 4  &\\
\hline $\tilde{\epsilon}$  & $\mathcal{U}(0.3, 3)$ & $0.62^{+0.19}_{-0.14}$ &  $ 0.48^{+0.11}_{-0.08}$ &  $0.54^{+0.10}_{-0.10}$  &$0.61^{+0.09}_{-0.07} $ &\\
\hline $n_{\rm sur}~(\rm fm^{-3})$ & $\mathcal{U}(0.17, 0.36)$ & $0.22^{+0.03}_{-0.03} $& $ 0.20^{+0.02}_{-0.02}$& $0.20^{+0.02}_{-0.02}$ & $0.18^{+0.01}_{-0.01}$   &\\
\hline  $R_{1.4} ~(\rm km)$ & $\ $ & $11.43^{+0.71}_{-0.73}$ &  $11.65^{+0.48}_{-0.62}$ &  $11.69^{+0.58}_{-0.51}$  & $12.16^{+0.26}_{-0.31}$ &\\
\hline  $\Lambda_{1.4} $ & $\ $ & $164.33^{+72.42}_{-50.81}$ &  $193.95^{+58.01}_{-54.29} $ &  $193.35^{+64.17}_{-46.92} $  & $212.36^{+53.71}_{-32.66}$ &\\
\hline  $R_{2.0} ~(\rm km)$ & $\ $ & $12.51^{+0.79}_{-0.86}$ &  $12.66^{+0.64}_{-0.66}$ &  $12.95^{+0.53}_{-0.71}$  & $13.38^{+0.27}_{-0.41}$ &\\
\hline  $M_{\rm TOV}(M_{\odot})$ & $\ $ & $3.52^{+0.17}_{-0.08}$ &  $3.49^{+0.19}_{-0.11}$ &  $3.51^{+0.17}_{-0.12}$  & $3.65^{+0.18}_{-0.16}$ &\\
\hline  $n_{\rm c}(n_0)_{M_{\rm TOV}}$ & $\ $ & $4.41^{+0.18}_{-0.63}$ &  $4.73^{+0.37}_{-0.48}$ &  $4.46^{+0.36}_{-0.20}$  & $3.94^{+0.26}_{-0.02}$  &\\
\end{tabular*}
\end{ruledtabular}
    \vspace{-0.4cm}
\label{table:prior_posteroor_para}
\end{table*}

\subsubsection{Constraints from gravitational wave events}
The tidal deformability inferred from gravitational wave detection of binary neutron star mergers have also led to a steady improvement in our understanding of the dense matter EOS. Here, we incorporate the constraints from the gravitational wave events GW170817~\cite{2017PhRvL.119p1101A,2018PhRvL.121p1101A} and GW190814~\cite{2020ApJ...896L..44A} reported by the LIGO Scientific Collaboration. Additionally, we investigate the possibility that the mass-gap secondary object ($M=2.59^{+0.08}_{-0.09}\msun$) in GW190814 potentially being a strangeon star. 

When treating the gravitational wave events, we fix the chirp mass $M_{\mathrm{c}}=$ $\left(M_1 M_2\right)^{3 / 5} /\left(M_1+M_2\right)^{1 / 5}$ to the median value $M_{\mathrm{cl}}=1.186 \mathrm{M}_{\odot}$ for GW170817. Ref.~\cite{2021ApJ...918L..29R} has shown that the small bandwidth of the chirp masses has almost no significant influence on the posterior distribution, contributing less than the sampling noise. We therefore fix the chirp mass, which is also beneficial in reducing the dimensionality of the parameter space and hence the computational cost. To speed up the convergence of our inference process, we transform the gravitational wave posterior distributions to include the two tidal deformabilities, chirp mass and mass ratio $q$, simultaneously reweighing such that the prior distribution on these parameters is uniform. The posterior then becomes
\begin{equation}
\begin{aligned}
& p(\boldsymbol{\theta}, \varepsilon_{\rm c} \mid \boldsymbol{d}, \mathcal{M}) \propto p(\boldsymbol{\theta} \mid \mathcal{M}) p(\varepsilon_{\rm c} \mid \boldsymbol{\theta}, \mathcal{M}) \\
& \times \prod_i p\left(\Lambda_{1, i}, \Lambda_{2, i}, q_i \mid \mathcal{M}_{\rm c}, \boldsymbol{d}_{\mathrm{GW}, \mathrm{i}}\left(, \boldsymbol{d}_{\mathrm{EM}, \mathrm{i}}\right)\right) \\
& \times \prod_j p\left(M_j, R_j \mid \boldsymbol{d}_{\mathrm{NICER}, \mathrm{j}}\right) \ ,
\end{aligned}
\end{equation}
where $\Lambda_{2, i}=\Lambda_{2, i}\left(\boldsymbol{\theta} ; q_i\right)$ is the tidal deformability, 
with the $i$th indicating the individual-event gravitational wave likelihood marginalized over all binary parameters. We follow the same convention as demonstrated in Ref.~\cite{2018PhRvL.121p1101A} and define $M_1>M_2$, since the gravitational wave likelihood function is degenerate under the exchange of the binary components.

All the inferences in this work employ the \texttt{CompactObject} package\cite{2024arXiv241114615H}, developed by the author and detailed in the Zenodo repository \cite{EoS_inference}. \texttt{CompactObject} is the inaugural open-source package that is extensively documented and offers comprehensive functionalities for applying Bayesian methods to constrain the EOS of dense matter. It supports various EOS models, such as relativistic mean field (RMF) and polytropes, and has been utilized in other studies such as \cite{2024MNRAS.529.4650H,2024arXiv241014572H}. For the Bayesian inference, we used the \texttt{UltraNest} package \cite{buchner2021ultranestrobustgeneral}, specifically employing its slice sampler, which is available on Github. We chose to use 50,000 live points for each inference to establish a baseline for comparing Bayes evidence, ensuring efficient and consistent high-dimensional sampling and convergence speeds.

\section{Results and discussions}\label{Sec: Results}

\subsection{Three-parameters case}
In Fig.~\ref{fig:posterior_NICER}, we present the posterior distributions of EOS parameters in three-parameter model, where we show the typical cases of $N_{\rm q}=9, 18, 27$ in each panel. 
The left panel depicts the joint analyses of PSR J0030+0451 and PSR J0740+6620, while the right panel includes additional observation data incorporating PSR J0437-4715. The contour levels in each corner plot indicate the 68.3\%, 95.4\%, and 99.7\% confidence regions, shaded from dark to light, respectively. Comparing the posterior parameter space across different choices of $N_{\rm q}$, we could explore the optimal choice for this quantity.
Our results demonstrate that increasing $N_{\rm q}$ in these joint Bayesian analyses favors larger values of $\epsilon$ and smaller values of $n_{\rm sur}$. This trend aligns with our theoretical understanding: as $N_{\rm q}$ increases, the EOS softens, leading to a deeper potential $\epsilon$ and a reduced surface baryon density $n_{\rm sur}$ to counteract the additional softening by enhancing the repulsive interactions. 
Comparing the results across two plots under the same $N_{\rm q}$, including PSR J0437-4715 results in a smaller $\epsilon$ and correspondingly smaller $n_{\rm sur}$. 
With other parameters unchanged, a smaller $\epsilon$ results in a softer EOS, and a smaller $n_{\rm sur}$ corresponds to a stiffer EOS. Therefore, the inclusion of this new observation has a mixed effect on the change in EOS parameters.

Bayes factors, defined as $\rm log$$ \ K = \log(Z_1 / Z_2)$, where $Z$ represents the Bayesian evidence, are employed to compare the effectiveness of models 1 and 2 in reconstructing the injected EOS. Per the standards in Ref.~\cite{Kass}, the model 1 is substantially preferred if its Bayes factor is above 3.2, strongly preferred when the factor exceeds 10, and decisive with a Bayes factor greater than 100.
Table~\ref{table:bayes_factor} presents the Bayesian evidence for several selected cases with varying $N_{\rm q}$ under the finite $N_{\rm q}$ value assumption motivated by the strangeon matter hypothesis. Under the constraints of PSR J0030+0451 and PSR J0740+6620, comparing the Bayesian evidence for $N_{\rm q}=18$, $N_{\rm q}=24$, and $N_{\rm q}=27$ models with $N_{\rm q}=9$ model yields the Bayes factors of $K =  Z_1/Z_2=99.5$, $244.7$, and $221.4$, respectively. The generally growing evidence indicates that the data predominantly favor the large $N_{\rm q}$ model. Comparing the $N_{\rm q}=24$ model with the $N_{\rm q}=18$ model results in a Bayes factor of $K = Z_1/Z_2=7.9$, while the comparison between the $N_{\rm q}=27$ and $N_{\rm q}=18$ models yields $6.3$. Notably, although increasing $N_{\rm q}$ from $18$ to $24$ improves the evidence from $-29.8$ to $-28.9$, these gains are relatively modest. This result also indicates a slight preference for models with larger large $N_{\rm q}$. However, the Bayes factor between $N_{\rm q}=27$ and $N_{\rm q}=24$ models is $K =  Z_1/Z_2=0.79$, which is not worth more than a bare mention and does not indicate a clear preference between these two models. Therefore, the analysis relatively supports the EOS model with $N_{\rm q} = 18$, assuming a finite value of $N_{\rm q}$. 
When additional data from PSR J0437-4715 are incorporated, comparisons of the larger $N_{\rm q}$ models with $N_{\rm q}=18,~24,~27$ to $N_{\rm q}=9$ yield the Bayes factors of $K = Z_1/Z_2=89~322$, $40~135$, and $17~9872$, respectively. The large Bayes factors reinforce the preference for models with larger $N_{\rm q}$ , particularly the $N_{\rm q}=18$ model. Table~\ref{table:bayes_factor} shows the Bayesian evidence reaches a local maximum at $N_{\rm q} = 18$ compared to $N_{\rm q} = 24$ and $N_{\rm q} = 9$ models. Increasing $N_{\rm q} = 18$ to $27$ results in a slight rise in evidence by about $0.7$, corresponding to a Bayes factor of $K=2.01$ between the $N_{\rm q}=27$ and $N_{\rm q}=18$ models, which is not statistically significant. Consequently, the preference for $N_{\rm q} = 18$ is further strengthened under the additional constraint from PSR J0437-4715, with an exceptionally huge Bayes factor of $K = 89~322$. The statistical evidence favors the $N_{\rm q}=18$ model for the strangeon matter, a strangeon consisting of $18$ quarks, which is completely symmetric in spin, flavor, and color spaces~\cite{1988PhRvL..60..677M,1991NuPhS..24...33C}. In this case, the number 18 corresponds exactly to the product of the internal degrees of freedom within the strangeon, calculated as $2 \times 3 \times 3 = 18$, representing spin, flavor, and color degrees of freedom, respectively. This provides a compelling physical motivation to select the $N_{\rm q} = 18$ for further study, with the assumption of finite $N_{\rm q}$ values.

Thus in the subsequent analyses, we fixed the number of quarks in one strangeon to $N_{\rm q} = 18$. 
Fig.~\ref{fig:compare_three_and_two_parameter} displays the marginalized posterior distribution functions for the EOS parameters, derived from four distinct cases of joint analyses under various astronomical constraints, described as follows:
\begin{itemize}
\item Case 1: The joint analysis of PSR J0030+0451 ($M=1.34_{-0.16}^{+0.15} \msun, R=12.71_{-1.19}^{+1.14} \mathrm{~km}$) and the heavy pulsar PSR J0740+6620 ($M=2.07 \pm 0.07 \msun, R=12.39_{-0.98}^{+1.30} \mathrm{~km}$). 
\item Case 2: Joint analysis including PSR J0030+0451, PSR J0740+6620, and PSR J0437-4715 ($\sim1.418 \msun, \sim11.36$~km). 
\item Case 3: Analysis under the combined constraints from PSR J0030+0451, PSR J0740+6620, PSR J0437-4715, and the GW170817 gravitational wave event.
\item Case 4: The most comprehensive case, including data from PSR J0030+0451, PSR J0740+6620, PSR J0437-4715, along with gravitational wave observations from GW170817, and the mass measurement of GW190814's secondary component, $2.59_{-0.09}^{+0.08}\msun$ (at the $90 \%$ confidence level) as a lower bound on the maximum mass.
\end{itemize}

Table~\ref{table:prior_posteroor_para} presents posterior values of the EOS parameters, along with their 68.3\% confidence intervals, and the most probable intervals of the strangeon star properties with 90\% confidence levels. The preferred parameter estimates for the Case 1 analysis are $\epsilon = 12.63^{+2.52}_{-1.87}\mev$ and $n_{\rm sur} = 0.23^{+0.03}_{-0.03}\fm3$. Incorporating the constraint from GW170817 results in a clear trend toward a slightly stiffer EOS by comparing the results from Case 2 and Case 3, as evidenced by an increase in $\epsilon$ from $11.02^{+1.61}_{-0.73}\mev$ to $11.76^{+1.97}_{-1.35}\mev$, while the posterior distributions for $n_{\rm sur}$ in Case 2 is the same as Case 3's results. Although including PSR J0437-4715 in Case 2 analysis further reduces $\epsilon$ compared with Case 1, thereby softening the EOS, it simultaneously introduces a stiffening effect by decreasing the surface density $n_{\rm sur}$. 
The parameter $\epsilon$ exhibits slight differences in its posterior distribution between these cases, reflecting subtle variations in the inferred EOS properties as additional constraints are considered. 
Interestingly, once $N_{\rm q}$ is fixed at 18, the normalized $\tilde{\epsilon}$ can be calculated. As the number of observations increases, this ratio remains approximately constant, around 0.65. Surprisingly, the inclusion of GW190814 does not further increase or decrease the normalized $\tilde{\epsilon}$ in Case 4 compared to previous cases in three-parameter EOS model. The secondary object of GW190814 is classified as a mass-gap object, which typically necessitates a much stiffer EOS, and consquently may cause a large chage in the corresponding EOS parameters. For example, the color-flavor-locked strange star should have a pairing gap larger than $244\mev$ to satisfy the constraint of GW190814's secondary component of $2.6\msun$ within the MIT bag model~\cite{2021ApJ...917L..22M}. However, the strangeon EOS employed in this study is inherently stiff, allowing it to easily satisfy the high-mass observational constraints. 
This demonstrates a significant advantage of the strangeon EOS, as it can simultaneously satisfy the very high mass criteria while adequately explaining all other observations. Additionally, the reason why including this mass-gap object did not substantially refine the posterior is that this object only provides mass information without accompanying radius or tidal deformability data. Consequently, the single-mass information on the secondary component of GW190814 results in looser constraints on the strangeon matter EOS.

\begin{figure*}
\centering 
{\includegraphics[width=0.49\textwidth]{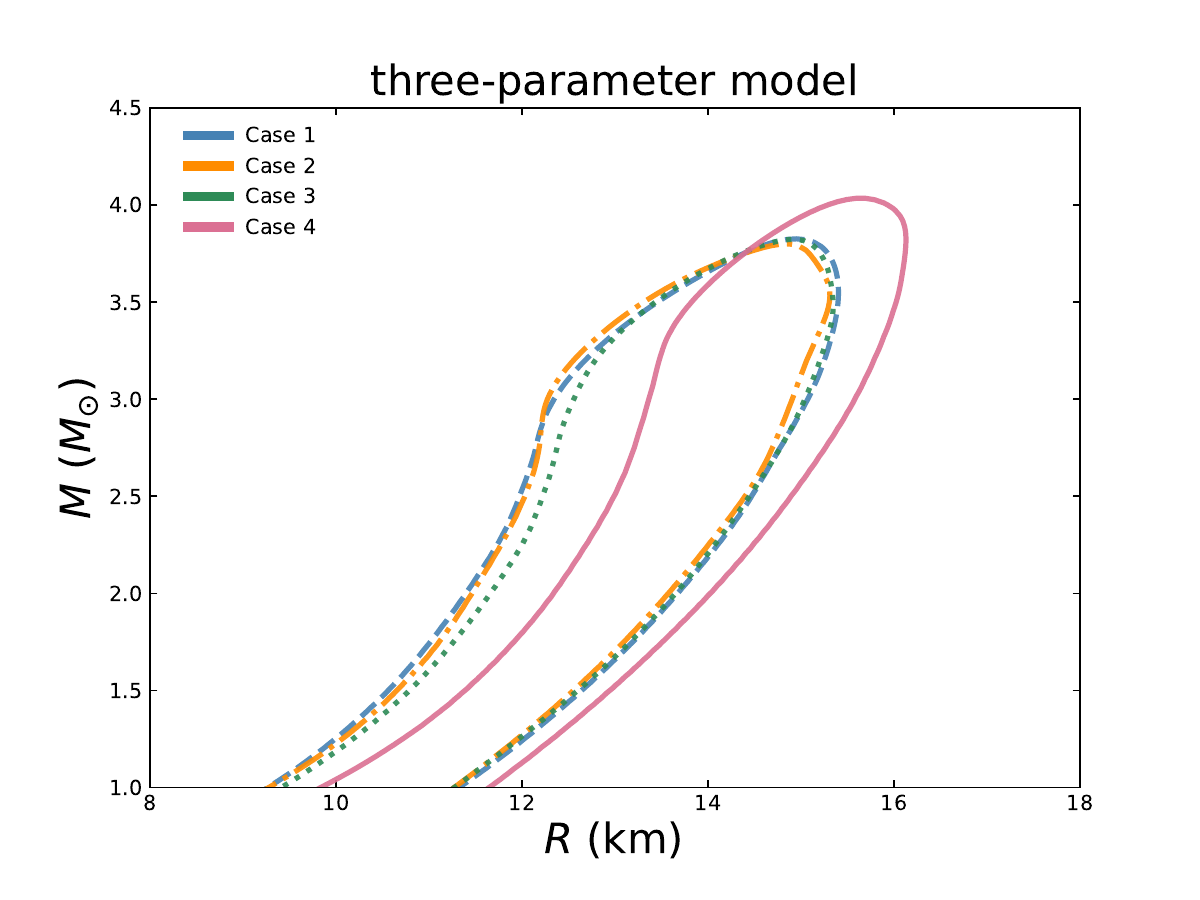}}
{\includegraphics[width=0.49\textwidth]{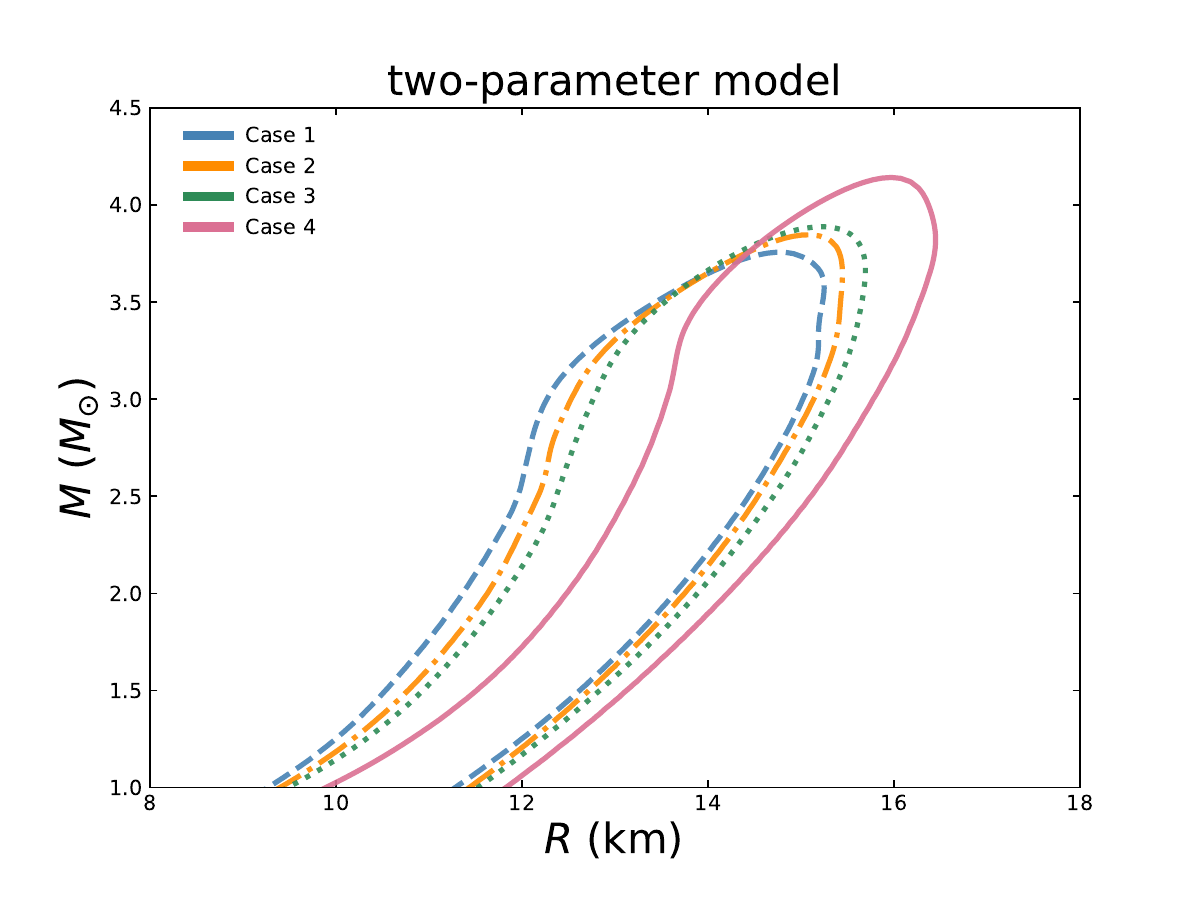}}
\caption{The M-R posterior distributions at $90\%$ confidence level resulting from the three-parameter and two-parameter models.} 
\label{fig:scatter_two_parameter}
\end{figure*}

\subsection{Comparison with two-parameter case}
As we mentioned before, the free parameters for strangeon matter EOS can be reduced to two by defining $\tilde{\epsilon}=\epsilon / N_{\rm q}$ and $\bar{n}=N_{\rm q} n / n_{\rm sur}$. The influence of $\epsilon$ and $N_{\rm q}$ on mass and radius is concurrent, thus only changing $\tilde{\epsilon}$ can produce a different M-R curve, that is to say, the parameters $\tilde{\epsilon}$ and $n_{\rm sur}$ fully determine the EOS stiffness and the shape of the M-R curve. An increase in the average potential depth per strangeon, $\tilde{\epsilon}$, and a reduction in surface baryon number density, $n_{\rm sur}$, results in a stiffer EOS due to the enlarged phenomenologically repulsive force. In this section, we aim to test the results between these models with different degrees of freedom EOS parameters. 

Like three-parameter model, the right panel of Fig.~\ref{fig:compare_three_and_two_parameter} displays the posterior distributions for the two-parameter EOS model, $\tilde{\epsilon}$ and $n_{\rm sur}$, while  Table~\ref{table:prior_posteroor_para} provides the posteriors of the EOS parameters and their 68.3$\%$ confidence range. 
As illustrated in Fig.~\ref{fig:compare_three_and_two_parameter}, a heavy star of mass $~2.1 \msun$ and a $~1.4 \msun$ star put constraints on the parameters, suggesting $\tilde{\epsilon} =0.62^{+0.19}_{-0.14}\mev$, $n_{\rm sur}=0.22^{+0.03}_{-0.03}\fm3$. When considering additional constraints from PSR J0437-4715, the results show a lower value for both $\tilde{\epsilon} = 0.48^{+0.11}_{-0.08}\mev$ and $n_{\rm sur}=0.20^{+0.02}_{-0.02}\fm3$. Therefore, it is hard to evaluate the direct influence of this constraint on the EOS property, since a lower $\tilde{\epsilon}$ leads to a softer EOS while a smaller $n_{\rm sur}$ results in a stiffer EOS, which is consistent with the results of the Bayesian analysis for the three-parameter model under the same constraints. However, the gravitational wave event GW170817 has a straightforward influence on $\tilde{\epsilon}$ by comparison of Case 2 and Case 3. GW170817 requires a slightly stiffer EOS to accommodate a slightly larger radius than PSR J0437-4715.
Previous studies have suggested that the secondary object in GW190814 could potentially be a quark star composed of the interacting quark matter~\cite{2021PhRvD.103f3018Z,2021PhRvL.126p2702B,2022PhRvD.106h3007C}. Given the inherently stiff nature of the strangeon matter EOS, we also consider the constraints of GW190814. The results obviously indicate that a stiffer EOS with larger $\tilde{\epsilon} =0.61^{+0.09}_{-0.07}\mev$ and smaller $n_{\rm sur}=0.18^{+0.01}_{-0.01}\fm3$ is required to support this observation. 

\subsection{The M-R posteriors and maximum mass of strangeon stars} 
Mapping the EOS posteriors to M-R space facilitates the understanding of how observational constraints influence the EOS and delineates the allowable regions in M-R space based on specific EOS models informed by various sets of observational data. We show in Fig.~\ref{fig:scatter_two_parameter} the M-R contours corresponding to the posterior distributions of the strangeon star EOSs. Every point in the EOS parameter space is uniquely correlated to a point in the EOS posterior parameter space. Then by varying central density, EOS points can be mapped to the M-R plane through integrating the Tolman-Oppenheimer-Volkoff (TOV) equations. 

Fig.~\ref{fig:scatter_two_parameter} illustrates that incorporating additional astronomical observations in the Bayesian analysis refines the M-R space for both three- and two-parameter models, yielding more constrained M-R relations. The inherent stiffness of the strangeon matter EOS leads to the dominant role of the PSR J0740+6620 compared to the results of Bayesian analyses in Cases 1, 2, and 3, with the addition of further observational constraints only minor changing the shape of the M-R curve. 
In particular, the inclusion of the GW190814 constraint excludes excessively soft EOSs, thus supporting the existence of superheavy compact stars around \( \sim 4.0 \, M_\odot \). This suggests that strangeon matter could feasibly explain the nature of massive compact objects that may be observed in the future. 
One can also evaluate the star's important properties as illustrated in Table~\ref{table:prior_posteroor_para} in detail. For instance, the radius of a $1.4\msun$ star is constrained to $11.47^{+0.60}_{-0.51}\rm \ km$ ($11.69^{+0.58}_{-0.51}\rm ~km$) within the three-parameter (two-parameter) model for Case 3. Similarly, the radius for a $2.0\msun$ star is approximately $1\rm km$ larger than that of a $1.4 \msun$ star, yielding values of $12.62^{+0.69}_{-0.60}\rm\  km$ ($12.95^{+0.53}_{-0.71}\rm\  km$) for the three-parameter (two-parameter) model, which aligns well with expectations, as the observed difference in radii between PSR J0437-4715 and PSR J0740+6620 is also approximately $1\rm\ km$. In these analyses, both three-parameter and two-parameter EOSs can easily support the existence of superheavy compact objects with masses exceeding $\sim 2.6\msun$. The maximum mass of strangeon stars can be $\sim 3.8\msun$. Incorporating the GW190814 constraint further shifts the results to favor even more massive compact stars, reinforcing the potential existence of extremely dense stellar configurations.

\section{Summary}\label{Sec: Summary}
In conclusion, we use Bayesian analysis to explore the parameter space of the EOSs for strangeon matter, constrained by recent astronomical observations. In particular, this analysis includes the new mass and radius measurements for PSR J0437-4715 and the secondary component of the gravitational wave event GW190814, a mass-gap object. Given the limited constraints from terrestrial experiments on strangeon matter, these astronomical observations play a crucial role in guiding the parameter space for this exotic matter. Our study also provides a comparative analysis of the posterior EOS parameter spaces derived from two different models: a three-parameter model and a two-parameter model. By subjecting both models to identical observational constraints, we assess their respective adaptability to the data. Despite differences in their theoretical formulations, the results from both models exhibit consistency, predicting a stiff EOS. This stiffness enables both models to accommodate the mass-gap object observed in GW190814 while still satisfying all current observations. This consistency aligns with the inherently stiff nature of the strangeon EOS model and demonstrates the advantage of this EOS model in explaining the observational data. The results suggest that current astronomical observations support $18$ quarks per strangeon, $N_{\rm q}=18$, indicating a preference for this quark configuration in the strangeon-matter model. This preference provides valuable insights into the underlying microphysical structure of strangeon matter and its implications for astrophysics.

When fixing $N_{\rm q}=18$, Bayesian analyses of both the two- and three-parameter models yield a consistent ratio of $\epsilon/N_{\rm q}$ around 0.6. Considering observational constraints from PSR J0030+0451, PSR J0740+6620, PSR J0437-4715, and GW170817, the inferred radius for a $1.4\msun$ star is $11.47^{+0.60}_{-0.51}~ \rm km$ in the three-parameter model, increasing slightly to $11.69^{+0.58}_{-0.51}~\rm km$ in the two-parameter model. The corresponding tidal deformability for a $1.4\msun$ star is $170.44^{+48.07}_{-40.62}$ and $193.35^{+64.17}_{-46.92}$ for three- and two-parameter models, with a narrow uncertainty for the $90\%$ confidence level, because of the stiff nature of strangeon matter. For a massive $2.0\msun$ star, the corresponding radii are $12.62^{+0.69}_{-0.60}~\rm km$ and $12.95^{+0.53}_{-0.71}~\rm km$, respectively. By incorporating the mass measurement of GW190814's secondary component, $2.59_{-0.09}^{+0.08} \msun$ (at the $90 \%$ confidence level), as a lower bound on the maximum mass, increases the upper boundary of $\tilde{\epsilon}$ from $0.64$ to $0.70$ in the two-parameter model at the $68.3\%$ confidence level. Due to the stiff nature of strangeon matter EOS, the change remains modest. In addition, we find that, for a $2.6~M_{\odot}$ star like GW190814's secondary component, the radius is found to be $14.33^{+0.29}_{-0.45}\rm km$. Future measurements of such massive compact objects could provide further insights into the dense matter EOS and the internal nature of compact objects.

Our results could be relevant to a hot topic of multiquark states~\cite{2016PhR...639....1C,2018RvMP...90a5004G}. What is the state of strongly interacting matter certainly remains a fundamental question directly related to the physics of compact stars. Extensive Efforts to resolve the mystery of dense matter have motivated several proposals for alternatives to neutron matter in the interior of compact stars, including strange quark matter~\cite{1998PhLB..438..123D,1989PhLB..229..112C,1993PhRvD..48.1409C,1999PhLB..457..261B,1999PhRvC..61a5201P,2005PhRvC..72a5204W,2010MNRAS.402.2715L,2018PhRvD..97h3015Z,2018PhRvD..98h3013L,2019PhRvD..99j3017X,2000PhRvC..62b5801P,2021EPJC...81..612B,2021ChPhC..45e5104X,2022PhRvD.105l3004Y,2024PhRvD.110j3012Z,2024FrASS..1109463Z}, two-flavor quark matter~\cite{2018PhRvL.120v2001H,2019PhRvD.100d3018Z,2020PhRvD.102h3003R,2020PhRvD.101d3003Z,2021PhRvD.103f3018Z,2024Ap&SS.369...29S}, color superconducting quark matter~\cite{2002PhRvD..66i4007S,2002JHEP...06..031A,2003PhRvD..67g4024A,2003PhRvD..67f5015H,2007PhRvD..76g4026M,2021PhRvD.103h3015R,2022PhRvD.105k4042I,2023PhRvD.108d3008Y,2024ApJ...966....3Y,2024arXiv241104064G}, quarkyonic matter with baryonic excitations near the Fermi surface~\cite{2007NuPhA.796...83M,2011PhLB..696...58H,2011PhRvD..84c4028S,2012NuPhA.877...70K,2020PhRvD.102b3021Z,2022PhRvD.105k4020C,2023PhRvD.108e4013X,2024arXiv241016649G}, and strangeon matter with strange clusters in position space~\cite{2003ApJ...596L..59X,2006MNRAS.373L..85X,2008MNRAS.384.1034P,2009MNRAS.398L..31L,2012MNRAS.424.2994L,2014MNRAS.443.2705Z,2016ChPhC..40i5102L,2018RAA....18...24L,2019EPJA...55...60L,2021RAA....21..250L,2022MNRAS.509.2758G,2022IJMPE..3150037M,2023PhRvD.108f3002Z,2023PhRvD.108l3031Z}, as an incomplete list of examples.
Among those efforts, the strangeon, as a kind of stable bound state,  could be natural for baryonic matter, which might attract particular interest in future studies of multi-quark states. Our Bayesian analysis strongly supports the scenario in which a strangeon forms a stable bound state with $N_{\rm q}=18$, exhibiting symmetry in color, flavor, and spin spaces. Investigating the interactions between strangeons provides valuable insights into strongly interacting matter and the EOS of dense matter. Nevertheless, advancing our understanding of strangeon matter further will rely on forthcoming experimental and observational developments, which are expected to provide crucial perspectives on its properties and its role in the composition of compact stars.

\medskip
\acknowledgments
We thank Prof. Lijing Shao and Prof. Kejia Lee for their valuable comments and helpful discussions. We also thank Zhiqiang Miao, Xiangdong Sun, Hanlin Song, and the PKU pulsar group for the helpful discussions and assistance with the use of supercomputers. This work is supported by the National SKA Program of China (2020SKA0120100), the National Natural Science Foundation of China (Nos. 12003047 and 12133003), the Strategic Priority Research Program of the Chinese Academy of Sciences (No. XDB0550300), and the Special Funds of the National Natural Science Foundation of China (Grant No. 12447171). C.H. acknowledges support from an Arts \& Sciences Fellowship at Washington University in St. Louis.


\begin{thebibliography}{0}%
\makeatletter
\providecommand \@ifxundefined [1]{%
 \@ifx{#1\undefined}
}%
\providecommand \@ifnum [1]{%
 \ifnum #1\expandafter \@firstoftwo
 \else \expandafter \@secondoftwo
 \fi
}%
\providecommand \@ifx [1]{%
 \ifx #1\expandafter \@firstoftwo
 \else \expandafter \@secondoftwo
 \fi
}%
\providecommand \natexlab [1]{#1}%
\providecommand \enquote  [1]{``#1''}%
\providecommand \bibnamefont  [1]{#1}%
\providecommand \bibfnamefont [1]{#1}%
\providecommand \citenamefont [1]{#1}%
\providecommand \href@noop [0]{\@secondoftwo}%
\providecommand \href [0]{\begingroup \@sanitize@url \@href}%
\providecommand \@href[1]{\@@startlink{#1}\@@href}%
\providecommand \@@href[1]{\endgroup#1\@@endlink}%
\providecommand \@sanitize@url [0]{\catcode `\\12\catcode `\$12\catcode `\&12\catcode `\#12\catcode `\^12\catcode `\_12\catcode `\%12\relax}%
\providecommand \@@startlink[1]{}%
\providecommand \@@endlink[0]{}%
\providecommand \url  [0]{\begingroup\@sanitize@url \@url }%
\providecommand \@url [1]{\endgroup\@href {#1}{\urlprefix }}%
\providecommand \urlprefix  [0]{URL }%
\providecommand \Eprint [0]{\href }%
\providecommand \doibase [0]{http://dx.doi.org/}%
\providecommand \selectlanguage [0]{\@gobble}%
\providecommand \bibinfo  [0]{\@secondoftwo}%
\providecommand \bibfield  [0]{\@secondoftwo}%
\providecommand \translation [1]{[#1]}%
\providecommand \BibitemOpen [0]{}%
\providecommand \bibitemStop [0]{}%
\providecommand \bibitemNoStop [0]{.\EOS\space}%
\providecommand \EOS [0]{\spacefactor3000\relax}%
\providecommand \BibitemShut  [1]{\csname bibitem#1\endcsname}%
\let\auto@bib@innerbib\@empty
\end{thebibliography}%


\begin{thebibliography}{99}
\bibliographystyle{apsrev4-1} 

\bibitem[Madsen(1999)]{1999LNP...516..162M} J. Madsen,\ 1999, Hadrons in Dense Matter and Hadrosynthesis, 162. 
\bibitem[Weber(2005)]{2005PrPNP..54..193W}  F. Weber,\ 2005, Prog. Part. Nucl. Phys., \textbf{54}, 193. 
\bibitem[Oertel et al.(2017)]{2017RvMP...89a5007O}M. Oertel, M. Hempel, T. Kl{\"a}hn, et al.\ 2017, Rev. Mod. Phys., \textbf{89}, 015007. 
\bibitem[Baym et al.(2018)]{2018RPPh...81e6902B}G. Baym, T. Hatsuda, T. Kojo, et al.\ 2018, Rep. Prog. Phys., \textbf{81}, 056902. 
\bibitem[Baiotti(2019)]{2019PrPNP.10903714B} L. Baiotti,\ 2019, Prog. Part. Nucl. Phys., \textbf{109}, 103714. 
\bibitem[Komoltsev \& Kurkela(2022)]{2022PhRvL.128t2701K} O.  Komoltsev, \& A. Kurkela, \ 2022, \prl, \textbf{128}, 202701. 
\bibitem[Miller et al.(2021)]{2021ApJ...918L..28M}M.~C. Miller, F.~K. Lamb, A.~J. Dittmann, et al.\ 2021, Astrophys. J. Lett., \textbf{918}, L28. 
\bibitem{2021ApJ...918L..27R} T.~E. Riley, et al., Astrophys. J. \textbf{918}, L27 (2021)
\bibitem{2017PhRvL.119p1101A} B.~P. Abbott, et al., Phys. Rev. Lett. \textbf{119}, 161101 (2017)
\bibitem{2018PhRvL.121p1101A} B.~P. Abbott, et al., Phys. Rev. Lett. \textbf{121}, 161101 (2018)
\bibitem{2019ApJ...887L..21R} T.~E. Riley, et al., Astrophys. J. \textbf{887}, L21 (2019)
\bibitem[Vinciguerra et al.(2024)]{2024ApJ...961...62V} S. Vinciguerra, T. Salmi, A.~L. Watts, et al.\ 2024, \apj, \textbf{961}, 62. 
\bibitem[Miller et al.(2019)]{2019ApJ...887L..24M}M.~C. Miller, F.~K. Lamb, A.~J. Dittmann, et al.\ 2019, Astrophys. J. Lett., \textbf{887}, L24. 
\bibitem[Choudhury et al.(2024)]{2024ApJ...971L..20C} D. Choudhury, T. Salmi, S. Vinciguerra, et al.\ 2024, Astrophys. J. Lett., \textbf{971}, L20. 
\bibitem[Rutherford et al.(2024)]{2024ApJ...971L..19R} N. Rutherford, M. Mendes, I. Svensson, et al.\ 2024, Astrophys. J. Lett., \textbf{971}, L19. 
\bibitem[Bodmer(1971)]{1971PhRvD...4.1601B} A.~R. Bodmer, \ 1971, \prd, \textbf{4}, 1601. 
\bibitem[Witten(1984)]{1984PhRvD..30..272W} E. Witten,\ 1984, \prd, \textbf{30}, 272. 
\bibitem[Chakrabarty(1993)]{1993PhRvD..48.1409C} S. Chakrabarty,\ 1993, \prd, \textbf{48}, 1409. 
\bibitem[Dey et al.(1998)]{1998PhLB..438..123D} M. Dey, I. Bombaci, J. Dey, et al.\ 1998, Phys. Lett. B, \textbf{438}, 123. 
\bibitem[Chakrabarty et al.(1989)]{1989PhLB..229..112C} S. Chakrabarty, S. Raha, \& B. Sinha, \ 1989, Phys. Lett. B, \textbf{229}, 112. 
\bibitem[Buballa \& Oertel(1999)]{1999PhLB..457..261B} M. Buballa, \&  M. Oertel,\ 1999, Phys. Lett. B, \textbf{457}, 261. 
\bibitem[Peng et al.(1999)]{1999PhRvC..61a5201P} G.~X. Peng, H.~C. Chiang, J.~J. Yang, et al.\ 1999, \prc, \textbf{61}, 015201. 
\bibitem[Wen et al.(2005)]{2005PhRvC..72a5204W} X.~J. Wen, X.~H. Zhong, G.~X. Peng, et al.\ 2005, \prc, \textbf{72}, 015204. 
\bibitem[Li et al.(2010)]{2010MNRAS.402.2715L} A. Li, R.~X. Xu, \& J.~F. Lu,\ 2010, Mon. Not. R. Astron. Soc., \textbf{402}, 2715. 
\bibitem[Zhou et al.(2018)]{2018PhRvD..97h3015Z}E.~P. Zhou, X. Zhou, \& A. Li,\ 2018, \prd, \textbf{97}, 083015. 
\bibitem[Li et al.(2018)]{2018PhRvD..98h3013L} C.~M. Li, Y. Yan, J.~J. Geng, et al.\ 2018, \prd, \textbf{98}, 083013. 
\bibitem[Xia et al.(2019)]{2019PhRvD..99j3017X} C.~J. Xia, T. Maruyama, N. Yasutake, et al.\ 2019, \prd, \textbf{99}, 103017. 
\bibitem[Peng et al.(2000)]{2000PhRvC..62b5801P} G.~X. Peng, H.~C. Chiang, B.~S. Zou, et al.\ 2000, \prc, \textbf{62}, 025801. 
\bibitem[Bai \& Liu(2021)]{2021EPJC...81..612B} Z. Bai,  \& Y.~X.  Liu.\ 2021, Eur. Phys. J. C, \textbf{81}, 612. 
\bibitem[Xia et al.(2021)]{2021ChPhC..45e5104X} C.~J. Xia, , Z. Zhu, X. Zhou, et al.\ 2021, Chinese Phys. C, \textbf{45}, 055104. 
\bibitem[Yuan et al.(2022)]{2022PhRvD.105l3004Y}W.~L. Yuan, A. Li, Z.~Q. Miao, et al.\ 2022, \prd, \textbf{105}, 123004. 
\bibitem[Zhou et al.(2024)]{2024PhRvD.110j3012Z} Y.~R.~Zhou, C. Zhang, J. Zhao, et al.\ 2024, \prd, \textbf{110}, 103012. 
\bibitem[Zhang et al.(2024)]{2024FrASS..1109463Z} X.~L. Zhang, Y.~F. Huang, \&  Z.~C. Zou,\ 2024, Front. Astron. Space Sci., 11, 1409463. 
\bibitem[Holdom et al.(2018)]{2018PhRvL.120v2001H}B. Holdom, J. Ren, \& C. Zhang,\ 2018, \prl, \textbf{120}, 222001. 
\bibitem[Zhao et al.(2019)]{2019PhRvD.100d3018Z} T. Zhao, W. Zheng, F. Wang, et al.\ 2019, \prd, \textbf{100}, 043018. 
\bibitem[Ren \& Zhang(2020)]{2020PhRvD.102h3003R}J. Ren,  \& C. Zhang, \ 2020, \prd, \textbf{102}, 083003. 
\bibitem[Zhang(2020)]{2020PhRvD.101d3003Z}C. Zhang, \ 2020, \prd, \textbf{101}, 043003. 
\bibitem[Zhang \& Mann(2021)]{2021PhRvD.103f3018Z}C. Zhang,  \& R.~B. Mann,\ 2021, \prd, \textbf{103}, 063018. 
\bibitem[Su et al.(2024)]{2024Ap&SS.369...29S} L.~Q.~Su, C.~Shi, Y.~F. Huang, et al.\ 2024, Astrophys.Space Sci., \textbf{369}, 29. 
\bibitem[Xu(2003)]{2003ApJ...596L..59X} R.~X. Xu, \ 2003, Astrophys. J. Lett., \textbf{596}, L59. 
\bibitem[Xu et al.(2006)]{2006MNRAS.373L..85X} R.~X. Xu, D.~J. Tao, \& Y. Yang, \ 2006, Mon. Not. R. Astron. Soc., \textbf{373}, L85. 
\bibitem[Peng \& Xu(2008)]{2008MNRAS.384.1034P}C. Peng,  \& R.~X. Xu, \ 2008, Mon. Not. R. Astron. Soc., \textbf{384}, 1034. 
\bibitem[Lai \& Xu(2009)]{2009MNRAS.398L..31L} X.~Y. Lai,  \& R.~X. Xu,\ 2009, Mon. Not. R. Astron. Soc., \textbf{398}, L31. 
\bibitem[Liu et al.(2012)]{2012MNRAS.424.2994L} X.~W. Liu, , J.~D. Liang, R.~X. Xu, et al.\ 2012, Mon. Not. R. Astron. Soc., \textbf{424}, 2994. 
\bibitem[Zhou et al.(2014)]{2014MNRAS.443.2705Z} E.~P. Zhou, J.~G. Lu, H. Tong, et al.\ 2014, Mon. Not. R. Astron. Soc., \textbf{443}, 2705. 
\bibitem[Lai \& Xu(2016)]{2016ChPhC..40i5102L} X.~Y. Lai,  \& R.~X. Xu, \ 2016, Chinese Phys. C, \textbf{40}, 095102. 
\bibitem[Lai et al.(2018)]{2018RAA....18...24L} X.~Y. Lai,  Y.~W. Yu, E.~P. Zhou, et al.\ 2018, Res. Astron. Astrophys., \textbf{18}, 024. 
\bibitem[Lai et al.(2019)]{2019EPJA...55...60L} X.~Y. Lai, E.~P. Zhou, \&R.~X. Xu, \ 2019,  Eur. Phys. J. A., \textbf{55}, 60. 
\bibitem[Lai et al.(2021)]{2021RAA....21..250L}X.~Y. Lai, C.~J. Xia, Y.~W. Yu, et al.\ 2021, Res. Astron. Astrophys., \textbf{21}, 250. 
\bibitem[Gao et al.(2022)]{2022MNRAS.509.2758G} Y. Gao, X.~Y. Lai, L.~J. Shao, et al.\ 2022, Mon. Not. R. Astron. Soc., \textbf{509}, 2758. 
\bibitem[Miao et al.(2022)]{2022IJMPE..3150037M}Z.~Q. Miao, C.~J. Xia, X.~Y. Lai, et al.\ 2022, Int. J. Mod. Phys. E, \textbf{31}, 2250037. 
\bibitem[Zhang et al.(2023)]{2023PhRvD.108f3002Z}C. Zhang, Y. Gao, C.~J. Xia, et al.\ 2023, \prd, \textbf{108}, 063002. 
\bibitem[Zhang et al.(2023)]{2023PhRvD.108l3031Z}C. Zhang, Y. Gao, C.~J. Xia, et al.\ 2023, \prd, \textbf{108}, 123031. 
\bibitem[Demorest et al.(2010)]{2010Natur.467.1081D} P.~B. Demorest, T. Pennucci, S.~M. Ransom, et al.\ 2010, \nat, \textbf{467}, 1081. 
\bibitem[Abbott et al.(2020)]{2020ApJ...896L..44A} R. Abbott, T.~D. Abbott, S. Abraham, et al.\ 2020, Astrophys. J. Lett., \textbf{896}, L44. 
\bibitem[Jones(1924)]{1924RSPSA.106..441J} J.~E. Jones, \ 1924,  Proc. R. soc. Lond. Ser. A, \textbf{106}, 441. 
\bibitem[Steiner et al.(2010)]{2010ApJ...722...33S} A.~W. Steiner, J.~M. Lattimer, \& E.~F. Brown,\ 2010, \apj, \textbf{722}, 33. 
\bibitem[Greif et al.(2019)]{2019MNRAS.485.5363G}S.~K. Greif, G. Raaijmakers, K. Hebeler, et al.\ 2019, Mon. Not. R. Astron. Soc., \textbf{485}, 5363. 
\bibitem[Raaijmakers et al.(2019)]{2019ApJ...887L..22R} G. Raaijmakers, T.~E. Riley, A.~L. Watts, et al.\ 2019, Astrophys. J. Lett., \textbf{887}, L22. 
\bibitem[Raaijmakers et al.(2020)]{2020ApJ...893L..21R} G. Raaijmakers, S.~K. Greif, T.~E. Riley, et al.\ 2020, Astrophys. J. Lett., \textbf{893}, L21. 
\bibitem[Traversi et al.(2020)]{2020ApJ...897..165T} S. Traversi, P. Char, \& G. Pagliara,\ 2020, \apj, \textbf{897}, 165. 
\bibitem[Xie \& Li(2021)]{2021PhRvC.103c5802X} W.~J. Xie,  \&B.~A. Li, \ 2021, \prc, \textbf{103}, 035802. 
\bibitem[Tak{\'a}tsy et al.(2023)]{2023PhRvD.108d3002T} J. Tak{\'a}tsy, P. Kov{\'a}cs, G. Wolf, et al.\ 2023, \prd, \textbf{108}, 043002. 
\bibitem[Huang et al.(2024)]{2024MNRAS.529.4650H} C. Huang, G. Raaijmakers, A.~L. Watts, et al.\ 2024, Mon. Not. R. Astron. Soc., 529, 4650. 
\bibitem[Marquez et al.(2024)]{2024PhRvD.110f3040M}K.~D. Marquez, T. Malik, H. Pais, et al.\ 2024, \prd, \textbf{110}, 063040. 
\bibitem[Guha Roy et al.(2024)]{2024PhLB..85939128G} Guha Roy, D., Venneti, A., Malik, T., et al.\ 2024, Phys. Lett. B, \textbf{859}, 139128. 
\bibitem[Huang et al.(2024)]{2024arXiv241014572H} C. Huang,  L. Tolos, C. Provid{\^e}ncia, et al.\ 2024, arXiv:2410.14572. 
\bibitem[Traversi \& Char(2020)]{2020ApJ...905....9T} S. Traversi, \& P. Char,\ 2020, \apj, \textbf{905}, 9. 
\bibitem[Miao et al.(2021)]{2021ApJ...917L..22M}Z.~Q. Miao, J.~L. Jiang, A. Li, et al.\ 2021, Astrophys. J. Lett., \textbf{917}, L22. 
\bibitem[Li et al.(2021)]{2021MNRAS.506.5916L} A. Li, Z.~Q. Miao, J.~L. Jiang, et al.\ 2021, Mon. Not. R. Astron. Soc., \textbf{506}, 5916. 
\bibitem[Wang et al.(2024)]{2024arXiv240911103W}Z. Wang, Y. Gao, D. Liang, J.~Zhang, \& L.~J.~Shao.\ 2024, arXiv:2409.11103. 
\bibitem[da Silva et al.(2024)]{2024PhRvD.109d3054D} F.~M. da Silva, A. Issifu, L.~L. Lopes, et al.\ 2024, \prd, \textbf{109}, 043054. 
\bibitem[Fonseca et al.(2021)]{2021ApJ...915L..12F} E. Fonseca, H.~T. Cromartie, T.~T. Pennucci, et al.\ 2021, Astrophys. J. Lett., 915, \textbf{L12}. 
\bibitem[Salmi et al.(2024)]{2024ApJ...974..294S}T. Salmi, D. Choudhury, Y. Kini, et al.\ 2024, \apj, \textbf{974}, 294. 
\bibitem[Bashinsky \& Jaffe(1997)]{1997NuPhA.625..167B} S.~V. Bashinsky, \& R.~L. Jaffe, \ 1997, Nucl. Phys. A, \textbf{625}, 167. 
\bibitem[Wetzorke et al.(2000)]{2000NuPhS..83..218W} I. Wetzorke, F. Karsch, \& E. Laermann,\ 2000, Nucl. Phys. B, Proc. Suppl., \textbf{83}, 218. 
\bibitem[Michel(1988)]{1988PhRvL..60..677M} F.~C. Michel, \ 1988, \prl, \textbf{60}, 677. 
\bibitem[Curtis Michel(1991)]{1991NuPhS..24...33C}F. Curtis Michel,\ 1991, Nucl. Phys. B, Proc. Suppl., \textbf{24}, 33. 
\bibitem[Stoks et al.(1994)]{1994PhRvC..49.2950S} V.~G.~J. Stoks, R.~A.~M. Klomp, C.~P.~F. Terheggen, et al.\ 1994, \prc, \textbf{49}, 2950. 
\bibitem[Wiringa et al.(1995)]{1995PhRvC..51...38W} R.~B. Wiringa, V.~G.~J. Stoks, \& R. Schiavilla,\ 1995, \prc, \textbf{51}, 38. 
\bibitem[Machleidt(2001)]{2001PhRvC..63b4001M}R. Machleidt, \ 2001, \prc, \textbf{63}, 024001. 
\bibitem[Reardon et al.(2024)]{2024ApJ...971L..18R} D.~J. Reardon,  M. Bailes, R.~M. Shannon, et al.\ 2024, Astrophys. J. Lett., \textbf{971}, L18. 
\bibitem[Raaijmakers et al.(2021)]{2021ApJ...918L..29R} G. Raaijmakers, S.~K. Greif, K. Hebeler, et al.\ 2021, Astrophys. J. Lett., \textbf{918}, L29. 
\bibitem[Huang et al.(2024)]{2024arXiv241114615H} C. Huang, T. Malik, J. Cartaxo, et al.\ 2024, arXiv:2411.14615. 
\bibitem[Huang et al.(2023)]{EoS_inference} C. Huang, G. Raaijmakers, A. L. Watts, L. Tolos, C. Provid\^{e}ncia, N. Osborn, \& N. Whitsett, \ 2023, GitHub Repository, version 1.9. 
\url{https://doi.org/10.5281/zenodo.10927600}; \url{https://github.com/ChunHuangPhy/EoS_inference/tree/v.1.9}
\bibitem[Buchner(2021)]{buchner2021ultranestrobustgeneral} J. Buchner,\ 2021, arXiv e-prints, arXiv:2101.09604. 
\url{https://johannesbuchner.github.io/UltraNest/}; \url{https://arxiv.org/abs/2101.09604} [stat.CO]
\bibitem[kass et al.(1995)]{Kass}
R.~E.~Kass and A.~E.~Raftery,
\emph{Bayes Factors}, 
\textit{Journal of the American Statistical Association} \textbf{90}, no.~430, pp. 773--795 (1995).
\bibitem[Bombaci et al.(2021)]{2021PhRvL.126p2702B} I. Bombaci, A. Drago, D. Logoteta, et al.\ 2021, \prl, 126, 162702. doi:10.1103/PhysRevLett.126.162702
\bibitem[Cao et al.(2022)]{2022PhRvD.106h3007C} Z. Cao, L.~W.~Chen, P.~C.~Chu , et al.\ 2022, \prd, \textbf{106}, 083007. 
\bibitem[Chen et al.(2016)]{2016PhR...639....1C}H. Chen, W. Chen, X. Liu, et al.\ 2016, Phys. Rep., \textbf{639}, 1-121. 
\bibitem[Guo et al.(2018)]{2018RvMP...90a5004G}F. Guo, C. Hanhart, U. Mei{\ss}ner, et al.\ 2018, Rev. Mod. Phys., \textbf{90}, 015004. 
\bibitem[Steiner et al.(2002)]{2002PhRvD..66i4007S} A.~W. Steiner, S. Reddy, \& M. Prakash,\ 2002, \prd, \textbf{66}, 094007. 
\bibitem[Alford \& Rajagopal(2002)]{2002JHEP...06..031A} M.~Alford, \& K.~Rajagopal, \ 2002, J. High Energy Phys., 2002, \textbf{031}. 
\bibitem[Alford \& Reddy(2003)]{2003PhRvD..67g4024A} M. Alford, \& S. Reddy, \ 2003, \prd, \textbf{67}, 074024. 
\bibitem[Huang et al.(2003)]{2003PhRvD..67f5015H} M. Huang, P. Zhuang, \& W. Chao, \ 2003, \prd, \textbf{67}, 065015. 
\bibitem[Mannarelli et al.(2007)]{2007PhRvD..76g4026M} M. Mannarelli, K. Rajagopal, \& Sharma, R.\ 2007, \prd, \textbf{76}, 074026. 
\bibitem[Roupas et al.(2021)]{2021PhRvD.103h3015R} Z. Roupas, G. Panotopoulos, \& I. Lopes, \ 2021, \prd, \textbf{103}, 083015. 
\bibitem[Ivanytskyi \& Blaschke(2022)]{2022PhRvD.105k4042I}  O. Ivanytskyi, \& D. Blaschke,\ 2022, \prd, \textbf{105}, 114042. 
\bibitem[Yuan et al.(2023)]{2023PhRvD.108d3008Y} W.~L. Yuan, J. Chao, \& A. Li, \ 2023, \prd, \textbf{108}, 043008. 
\bibitem[Yuan \& Li(2024)]{2024ApJ...966....3Y} W.~L. Yuan,\&A. Li,\ 2024, \apj, \textbf{966}, 3. 
\bibitem[Gholami et al.(2024)]{2024arXiv241104064G} H. Gholami, I.~A. Rather, M. Hofmann, et al.\ 2024, arXiv:2411.04064. 
\bibitem[McLerran \& Pisarski(2007)]{2007NuPhA.796...83M} L. McLerran, \&R.~D. Pisarski,\ 2007, Nucl. Phys. A, \textbf{796}, 83. 
\bibitem[Herbst et al.(2011)]{2011PhLB..696...58H} T.~K. Herbst, J.~M. Pawlowski, \& B.-J. Schaefer,\ 2011, Phys. Lett. B, \textbf{696}, 58. 
\bibitem[Shao et al.(2011)]{2011PhRvD..84c4028S} G.~Y. Shao, M. di Toro, V. Greco, et al.\ 2011, \prd, \textbf{84}, 034028. 
\bibitem[Kojo(2012)]{2012NuPhA.877...70K} T. Kojo,\ 2012, Nucl. Phys. A, \textbf{877}, 70. 
\bibitem[Zhao \& Lattimer(2020)]{2020PhRvD.102b3021Z} T.~Q.~Zhao,  \&J.~M. Lattimer, \ 2020, \prd, \textbf{102}, 023021. 
\bibitem[Cao(2022)]{2022PhRvD.105k4020C} G.~Q.~Cao, \ 2022, \prd, \textbf{105}, 114020. 
\bibitem[Xia et al.(2023)]{2023PhRvD.108e4013X} C.~J.~Xia, H.~M. Jin, \& T.~T. Sun, \ 2023, \prd, \textbf{108}, 054013. 
\bibitem[Gao \& Harada(2024)]{2024arXiv241016649G} B.~K.~Gao,  \& M. Harada, \ 2024, arXiv:2410.16649. 

\end{thebibliography}
\end{document}